# 1. Chiral two-dimensional $MoS_2$ by molecular functionalization as ultra-sensitive detectors for circularly polarized light

*Ye Wang\*, Yiru Zhu, Han Yan, Yang Li, Yan Wang, Manish Chhowalla\**

Department of Materials Science & Metallurgy, University of Cambridge, 27 Charles Babbage Road, Cambridge CB3 0FS, United Kingdom

\*M.C.: email, mc209@cam.ac.uk

\*Ye W.: email, yw596@cam.ac.uk

**Keywords**: Two-dimensional materials, transition metal dichalcogenides, chirality, circularly polarized light, circular dichroism

**ABSTRACT**

Inducing chirality in optically and electronically active materials is interesting for applications in sensing and quantum information transmission. Two-dimensional (2D) transition metal chalcogenides (TMDs) possess excellent electronic and optical properties but are achiral. Here we demonstrate chirality induction in atomically thin layers of 2D $MoS_2$ by functionalization with chiral thiol molecules L-/D-penicillamine (L-/D-PEN). Analysis of X-ray absorption near-edge structure and Raman optical activity with circularly polarized excitation suggest chemical and electronic interactions that leads chirality transfer from the molecules to the $MoS_2$. We confirm chirality induction in 2D $MoS_2$/L-PEN with circular dichroism measurements that show absorption bands at wavelengths of 380-520 nm and 520-600 nm with giant molar ellipticity of $10^8$ deg·cm$^2$/dmol – 2-3 orders of magnitude higher than 3D chiral materials. Phototransistors fabricated from atomically thin $MoS_2$/PEN for detection of circularly polarized light exhibit responsivity of $>10^2$ A/W and maximum anisotropy g-factor of 1.98 – close to the theoretical maximum of 2.0, which indicates that the chiral states of photons are fully distinguishable by the photodetectors. Our results demonstrate that it is possible achieve chirality induction in monolayer $MoS_2$ by molecular functionalization and realise ultra-sensitive detectors for circularly polarized photons.

# MAIN

## 1. Introduction

Chirality describes the absence of mirror symmetry.[1,2] Chiral materials and structures can be used as sensors for stereoselective recognition[3] and chemical processes for enantioselective catalysis[4]. Chiral states of photons or electrons can be used for information transmission in quantum networks.[5-7] Solid-state chiral materials with technologically interesting optical and electrical properties are just beginning to emerge. Recent work has shown that it is possible to transfer chirality from chiral organic molecules onto inorganic materials.[8-11] This provides a pathway for realising chiral materials that are optically and electronically active so that high performance devices such as photo-emitters[12-15] and photodetectors[16-19] for emission and detection of circularly polarised light can be realized.

In this work, we demonstrate chirality induction in two-dimensional (2D) transition metal dichalcogenides (TMDs) by functionalising them with chiral molecules. Monolayer 2D TMDs are direct band gap semiconductors with excellent opto-electronic properties.[20,21] They exhibit strong photoluminescence[22,23] and under special circumstances they can emit single photons[24,25]. Electronically, they are excellent channel materials for field effect transistors showing high mobility, high on/off ratio, and low subthreshold swing.[26-29] However, monolayer TMDs belong to $D_{3h}$ symmetry point group which contains a $S_3$ improper axis and therefore are geometrically and optically achiral.[30,31] Previous work has demonstrated that three-dimensional (3D) materials such as polymers, magnetic nanoparticles, hybrid perovskites, and TMDs that are naturally achiral can be made chiral by surface functionalization, interlayer intercalation, or induced self-assembly of chiral organic molecules.[8,10,17,19,32-35] The interactions between organic chiral molecules and inorganic materials can be weak[8,32,36-38] or strong.[9,39-41] The latter involves orbital hybridization which can result in new spin-polarized electronic states at energies different from the band edges of the inorganic material and the chiral organic molecules.[10,42]

Covalent functionalization of TMDs can be achieved by nucleophilic substitution with diazonium salts, organic halides, or maleimide derivatives.[43-47] Alternatively, organic thiols have been used to passivate sulphur vacancies in $MoS_2$ [48-54], which is less aggressive than nucleophilic reactions and therefore provide opportunity for inducing new functionalities in 2D TMDs while preserving their opto-electronic properties.

Here we use natural chiral amino-acids containing thiol group to functionalize 2D TMDs [**Figure 1(a)**]. Specifically, we have selected L-/D-penicillamine (L-/D-PEN) as the chiral molecules for functionalization of chemical vapor deposition (CVD) grown monolayer $MoS_2$. We use gentle evaporation of molecules as the method for thiol functionalization.

## 2. Results and Discussions

Functionalization of $MoS_2$ by L-/D-PEN was initially characterized by X-ray photoelectron spectroscopy (XPS). As shown in **Figure 1(b)**, core-level S 2p doublet of $MoS_2$ appears at 162.94/164.18 eV. After functionalizing with L-PEN, we observed an additional doublet peak at lower binding energy (162.19/163.42 eV, marked in red colour). This doublet is typical of bound thiol and is attributed to reduced sulphur in thiol that is chemically bonded with metal atoms. [55-57] We have not found evidence of free PEN molecules that possess S 2p peaks at 163.37/164.45 eV (inset of **Figure 1(b)**), indicating that weakly bound free thiol molecules are largely absent in our samples. The degree of chemisorbed functionalization estimated from XPS measurements is 14.47% for $MoS_2$/L-PEN and 12.02% for $MoS_2$/D-PEN. Additional XPS spectra of Mo 3d, N1s and C 1s, atomic force microscopy (AFM) and Fourier-transform infrared (FTIR) spectroscopy (**Figure S1-S3** in Supplementary Information) suggest that while signature functional groups of the molecule such as ($-NH_2$, -COOH) can be probed on $MoS_2$/L (D)-PEN, molecules do not form aggregates or islands on $MoS_2$ as indicated by the negligible (<0.1 nm) increase in roughness measured by AFM. Moreover, FTIR spectra show absence of thiol group (-SH) at 2598 $cm^{-1}$ in $MoS_2$/PEN, indicating conjugation of thiol molecules on $MoS_2$ (**Figure S3** in Supplementary Information).[56] To probe the subtle changes induced by functionalization, we used synchrotron-based X-ray absorption near-edge structure (XANES) to obtain atomic-orbital-specific information before and after molecular functionalization of monolayer $MoS_2$. **Figure 1(c)** shows the XANES Mo $L_3$-edge ($2p_{3/2} \rightarrow 4d$) of $MoS_2$[48,58]. No obvious shift or appearance of shoulder at higher energy that would indicate the presence of oxidized species such as $MoO_2$ (2524.5 eV) and $MoO_3$ (2528.6 eV) were observed post-functionalization.[59,48] The $d(I)/d(E)$ in **Figure 1(c)** shows a maximum at 2516.7 eV corresponding to the $Mo^{4+}$ absorption edge of both pristine and functionalized $MoS_2$. However, increased absorption as indicated by the downward arrow in **Figure 1(c)** at energies below the white line in functionalized samples was observed – indicating that the Mo orbitals in $MoS_2$ are altered by PEN functionalization.[60]

The sulphur K-edge (1s→3p) spectra are more sensitive to change in chemical environment of $MoS_2$. **Figure 1(d)** shows that monolayer $MoS_2$ exhibits four absorption bands, labelled as A, B, C and D with peak maxima at 2465.2 eV, 2467.3 eV, 2471.9 eV and 2476.5 eV, respectively. The absorption edge according to the first derivative in **Figure 1(d)** is defined at 2464.9 eV. Comparing to pristine $MoS_2$, L-/D-PEN functionalized $MoS_2$ displays three prominent features: 1) in $MoS_2$/L(D)-PEN, the intensity ratio between A and B bands decreases (more evident comparison can be found in the normalized spectra in **Figure S4** in the Supplementary Information); 2) band C redshifts 0.7 eV (from 2472.0 eV to 2471.3 eV); and 3) similar to Mo $L_3$-edge spectra, a weak pre-edge absorption occurs from 2463.5 eV to 2464.5 eV after molecular functionalization. A and B bands have been attributed to absorption due to excitation of sulphur 1$s$ core electrons excitation to $p_{x,y}$ and $p_z$ states, respectively.[61] Band C is a broad merged double band corresponding to transition from S 1$s$ electrons to $p_z$- (lower energy) and $p_{x,y}$- (higher energy) like states in the continuum, originating from hybridization with S 3$d$ orbitals.[58,62] We can therefore translate observations of decrease in intensity ratio between A and B bands and redshift of 0.68 eV to more 1s to 3$p_z$ transitions following molecular functionalization, which may be attributed to availability of more S $p_z$ states provided by thiol groups from PEN molecules on $MoS_2$. In all samples, we did not find absorption bands from free PEN molecules at 2473.1 eV,[63] indicating the changes in XANES are from sulphur orbitals modified by the molecule in $MoS_2$. The XANES results suggest that evaporation of chiral PEN molecules onto CVD grown $MoS_2$ leads to subtle but measurable orbital hybridization between the two.

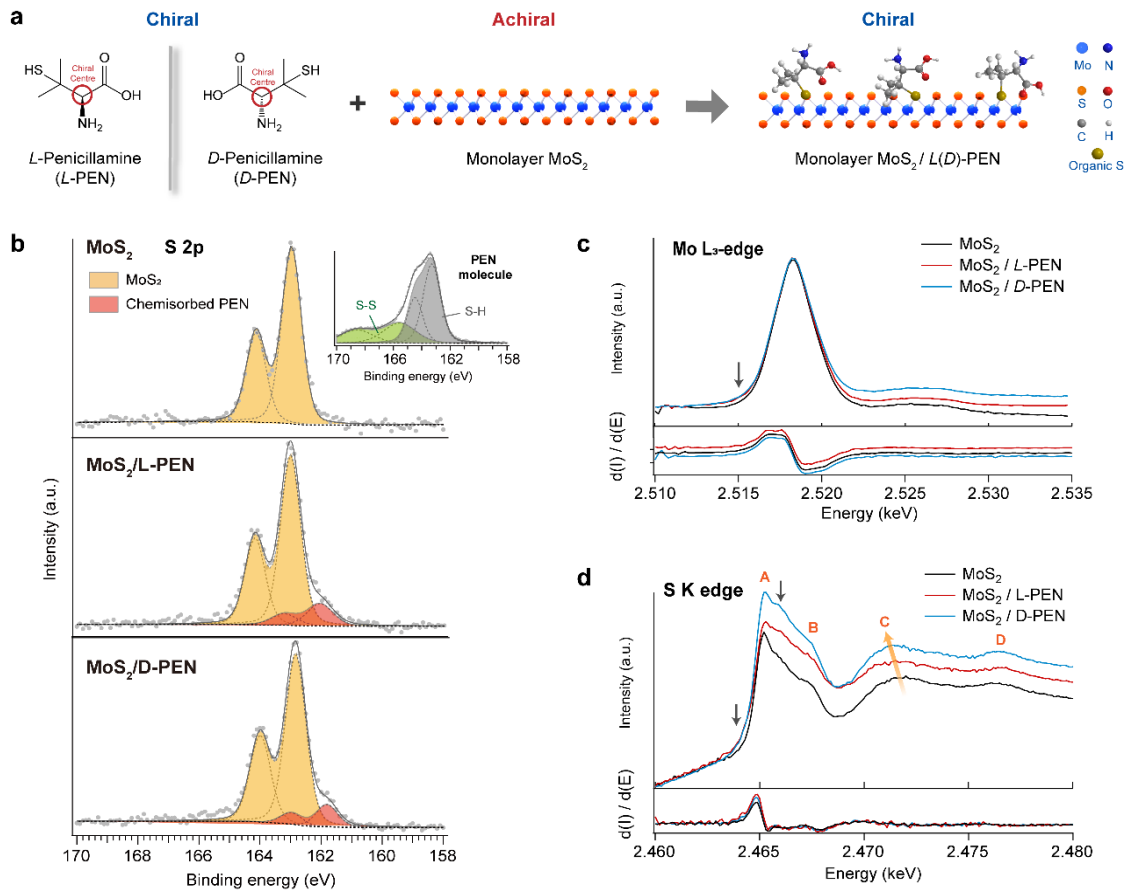

**Figure 1. (a)** Chemical structure of chiral molecules, L/D-Penicillamine (L/D-PEN) and schematic of monolayer MoS₂. The schematic represents the general scheme for surface functionalization with chiral thiol molecules. Sulphur (marked in yellow ball ) from thiol group in PEN molecules is primarily attached at sulphur vacancies of MoS₂ resulting in chemisorbed functionalization. **(b)** Core-level XPS S 2p spectra of MoS₂, MoS₂/L-PEN and MoS₂/D-PEN. In MoS₂/L-PEN and MoS₂/D-PEN the spectra show an additional component at lower binding energy (marked in red) than pure MoS₂. The inset shows the S 2p spectrum of pure L-PEN molecules. **(c-d)** (c) Mo L₃-edge and (d) S K-edge spectra of MoS₂, MoS₂/L-PEN and MoS₂/D-PEN. In (c) the absorption edge is shown on top and its derivative d(I)/d(E) is shown below. The downward arrow at the edge shows that the absorption of functionalized samples is higher, which suggests change in core electron bonding energy due to functionalization. In (d), the XANES of S edge is shown with the small downward arrow indicating that functionalized samples absorb more. The d(I)/d(E) curves for the three samples do not show significant differences. A, B, C, and D represent absorption bands from sulphur. The orange arrow shows that the C band disperses to lower energy.

Next, we assess the chiroptical properties of functionalized 2D MoS₂ by measuring circular dichroism (CD) spectra at room temperature. We use CVD-grown film transferred on quartz (surface coverage of ~ 80%), as shown in optical microscope images in **Figure 2(a)**. CD ($\Delta A$) is the difference in absorbance of left circularly polarised light (LCP) ( $A_{LCP}$) and right circularly polarized light (RCP) ( $A_{RCP}$):

$$\Delta A = A_{LCP} - A_{RCP} \qquad (1)$$

Chirality is related to the molar ellipticity ($[\theta]$):

$$[\theta] = \Delta\varepsilon \cdot 3298.2 \qquad (2)$$

where $\Delta\varepsilon$ is the molar CD which is the difference between the molar extinction coefficient of LCP and RCP. Molar ellipticity is generally used to compare chirality values that are measured by different experimental setups. Calculation of molar ellipticity involves Beer-Lambert law, the details of which are described in **Methods**.

The CD spectra (in the form of molar ellipticity versus wavelength) of pristine $MoS_2$ along with L(D)-PEN functionalized $MoS_2$ are shown in **Figure 2(b)**. The CD spectra of L(D)-PEN only are shown in the inset of **Figure 2(b)**. It can be seen that pristine $MoS_2$ is chiroptically silent while $MoS_2$/L(D)-PEN samples exhibit molar ellipticity of up to $10^8$ deg·cm$^2$/dmol. The ellipticity values of the functionalised 2D $MoS_2$ samples are 2-3 orders of magnitude higher than 3D chiral materials.[8,10,19,32] The CD bands of functionalized $MoS_2$ are located at ~ 380-520 nm and 520-600 nm. These are substantially different from the CD band of molecular L(D)-PEN that are located at 210-270 nm (inset of **Figure 2(b)**). The 380-520 nm band in functionalised $MoS_2$ is close to the C and D exciton peaks of $MoS_2$ as shown in **Figure S5** of Supplementary Information.[64] This may suggest that functionalization of $MoS_2$ with chiral molecules may induce interactions between the molecules and excitonic electronic states in $MoS_2$ that leads to chirality transfer. It can be seen in **Figure 2(b)** that in monolayer $MoS_2$ both L- and D-PEN/$MoS_2$ exhibit similar trend (both positive or both negative) in CD bands in **Figure 2(b).** This is in contrast with three dimensional (3D) chiral nanomaterials that show the sign conversion in left and right handed molecules – called the Cotton effect.[10,17,19,34,65]. That is, the CD bands of L- and D-PEN/$MoS_2$ should exhibit opposite signs (one is positive and other in negative) as shown in **Figure S6** of Supplementary Information. Indeed, few-layered $MoS_2$ obtained by liquid exfoliation and functionalized with chiral molecules also show sign inversion of Cotton effect (**Figure S6** of Supplementary Information and Ref [62]). Our results suggest that the mechanism for chirality induction and molecular functionalization of 2D materials such as $MoS_2$ is fundamentally different from that for 3D materials.

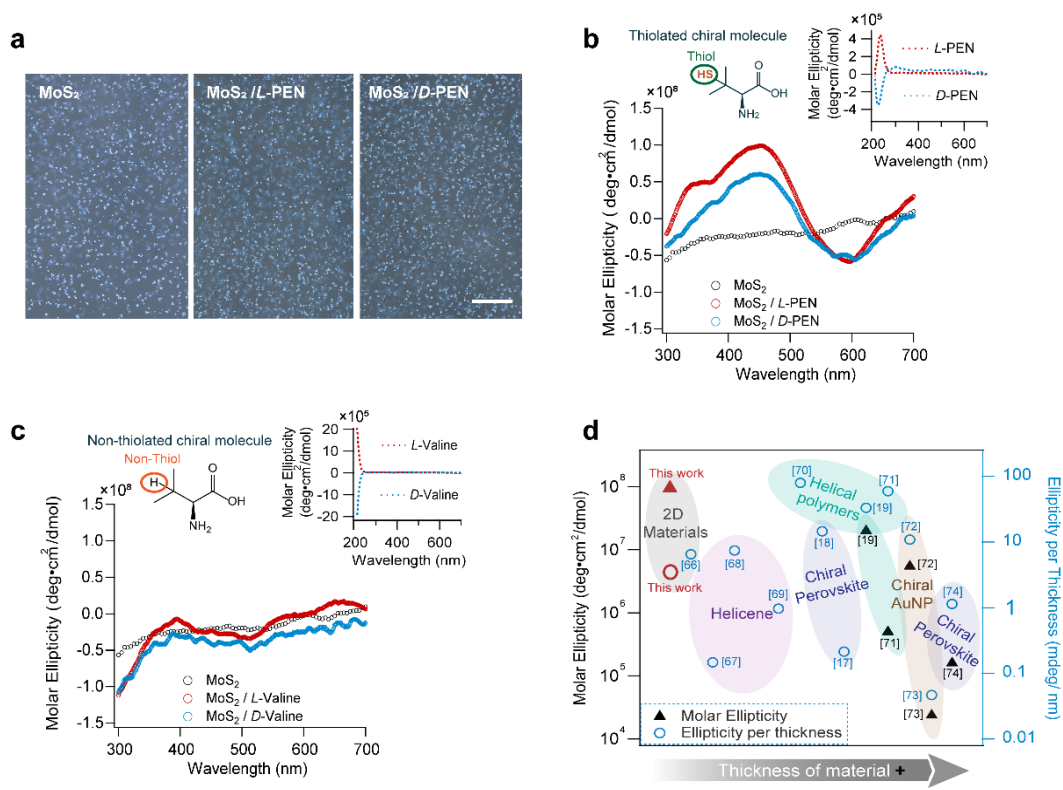

**Figure 2. (a)** Optical microscope images of CVD-grown film of MoS$_2$, MoS$_2$/ L-PEN and MoS$_2$/ D-PEN transferred on quartz substrates. The scale bar is 300 μm. **(b)** Molar ellipticity of MoS$_2$, MoS$_2$/ L-PEN and MoS$_2$/ D-PEN as a function of the wavelength. Inset shows CD of molecules only. The broad bands from functionalized MoS$_2$ located at ~ 380-520 nm and 520-600 nm are absent in pure molecule spectra. **(c)** Molar ellipticity of MoS$_2$, MoS$_2$/ L-Valine and MoS$_2$/ D-Valine showing absence of chirality induction. Valine is not thiolated and therefore does not covalently functionalize with MoS$_2$. Inset is for pure valine molecules. **(d)** Comparison of molar ellipticity and ellipticity per thickness with reported values.[17-19,66-74]

We perform reference experiments with non-thiolated chiral molecule, L/D-valine, to study the role of the thiol group of L(D)-PEN in imparting chirality to MoS$_2$ through chemical interactions. Chemical structure of valine is identical to that of penicillamine with the difference that the head group is an H atom instead of -SH. In the absence of the thiol group, we expect L(D) – valine molecules to be physisorbed on the surface of MoS$_2$ and therefore only weakly perturb the electronic structure or transfer chirality. The CD spectra of MoS$_2$ with L(D) – valine molecules are shown in **Figure 2(c)**. The CD spectra in **Figure 2(c)** do not show additional CD bands from MoS$_2$/L(D)-valine samples even though XPS results clearly show the presence of the molecules on MoS$_2$ after functionalization (**Figure S7** in the **Supplementary Information**). These results therefore suggest that chemical interactions of sulphur in L(D)-PEN with MoS$_2$ influences adjacent chiral carbon atoms, which leads to structural chirality transfer to MoS$_2$. To put our results in context, we compare them with those reported in the literature for solid-state 3D chiral materials in **Figure 2(d)**. It can be seen that the absolute molar ellipticity (solid triangles in **Figure 2(d)**) of L(D)-PEN functionalised $MoS_2$ is among the highest of any reported material. The molar ellipticity normalised to the thickness is comparable to other materials (hollow circles in **Figure 2(d)**), reaching 3.23 mdeg/nm.

The giant molar ellipticity in $MoS_2$/ PEN suggests strong dissymmetry in light-matter interactions. To further probe how hybridization between PEN and $MoS_2$ affects the lattice symmetry and chiroptical properties, we used linearly and circularly polarized Raman spectroscopy.[75] Since surface functionalization can lead to change in carrier density that can influence Raman signal, we first used linearly polarized light to eliminate doping of $MoS_2$ as the possible cause of changes in Raman. Monolayer $MoS_2$ show two typical $E^1_{2g}$ in-plane and $A_{1g}$ out-of-plane Raman vibrational modes [**Figure 3(a,b)**] under 532 nm linear laser excitation. After functionalization with L(D)-PEN, both peaks exhibit broadening and shifts to lower energy as shown in **Figure 3(c)**. We observe statistically insignificant redshift of $E^1_{2g}$ mode of $0.20\pm0.23$ cm$^{-1}$ in $MoS_2$/L-PEN and $0.33\pm0.31$ cm$^{-1}$ in $MoS_2$/D-PEN. A significant redshift of $0.87\pm0.05$ cm$^{-1}$ is found in $A_{1g}$ mode of $MoS_2$/L-PEN and $1.15\pm0.22$ cm$^{-1}$ in $MoS_2$/D-PEN. We also compare the full width half maximum (FWHM) of the peaks. We observe negligible change in FWHM of $E^1_{2g}$ peak while the $A_{1g}$ mode FWHM increases from $2.84\pm0.14$ cm$^{-1}$ in the pristine $MoS_2$ to $3.36\pm0.08$ cm$^{-1}$ in $MoS_2$/L-PEN and $4.43\pm0.06$ cm$^{-1}$ in $MoS_2$/D-PEN [**Figure 3(d)**]. Since the $E^1_{2g}$ mode is related to strain due to lattice distortions and $A_{1g}$ mode is affected by electron-phonon coupling caused by doping,[76-78] the results suggest that functionalized PEN molecules create mostly electronic perturbations in $MoS_2$. To evaluate whether these perturbations are chiral, we used circularly polarized excitation in Raman to access symmetry dependent helicity-resolved phonon scattering. Previous work has reported that pristine 2D TMDs are sensitive to circularly polarized photons, resulting in helicity-resolved Raman modes.[79-81] The circularly polarized Raman spectra of $MoS_2$, $MoS_2$/L-PEN and $MoS_2$/D-PEN are plotted in **Figure 3(e, i-iii)**. In both pristine and functionalized $MoS_2$, we observe that the peak intensity ratios of $E^1_{2g}$ and $A_{1g}$ ($I_{E^1_{2g}}/I_{A_{1g}}$) are generally higher for excitation by right circularly polarized photons (σ+) than for left circularly polarized excitation (σ−). The ratio values are listed in **Table S1** in **Supplementary Information**. We observe a significant rise in $I_{E^1_{2g}}/I_{A_{1g}}$ in chiral $MoS_2$/L-PEN from 0.865 to 0.988 with σ+ excitation while σ− excitation remains at similar value. In contrast, MoS$_2$/ D-PEN shows higher $I_{E^1_{2g}}/I_{A_{1g}}$ with σ− excitation while σ+ excitation is unchanged. The Raman optical activity (ROA) is given by:

$$ROA = I_R - I_L \qquad (3)$$

where $I_R$ and $I_L$ are the intensities of the MoS$_2$ Raman peaks when excited by right (σ+) and left (σ−) circularly polarized light, respectively. The resultant ROA spectra from the three samples plotted in **Figure 3(f)** display negative ROA of –83.4 in E$^1_{2g}$ and positive ROA of 140.7 in A$_{1g}$ vibrational modes of pristine MoS$_2$. The ROA in MoS$_2$/L-PEN shows an enhancement in the E$^1_{2g}$ to –114.4 and A$_{1g}$ to 192.2. For MoS$_2$/D-PEN, the E$^1_{2g}$ ROA is nearly tripled to –188.6 whereas an inversion in the sign of A$_{1g}$ ROA of –98.6 is observed. Considering that the measurements are taken on a single flake of monolayer MoS$_2$ with laser excitation spot size of ~ 2 μm, the Raman signals are intrinsically weak. However, ROA intensity of up to 21% of the Raman signal for MoS$_2$ indicates high dissymmetry in these phonon modes. The ROA results suggest that functionalization leads chirality transfer to the monolayer MoS$_2$ – influencing its vibrational modes [34]. Together with CD measurements in **Figure 2**, the chemical functionalization of 2D MoS$_2$ with chiral thiol molecules induces high order of structural chirality that can interact with circularly polarized light with high efficiency.

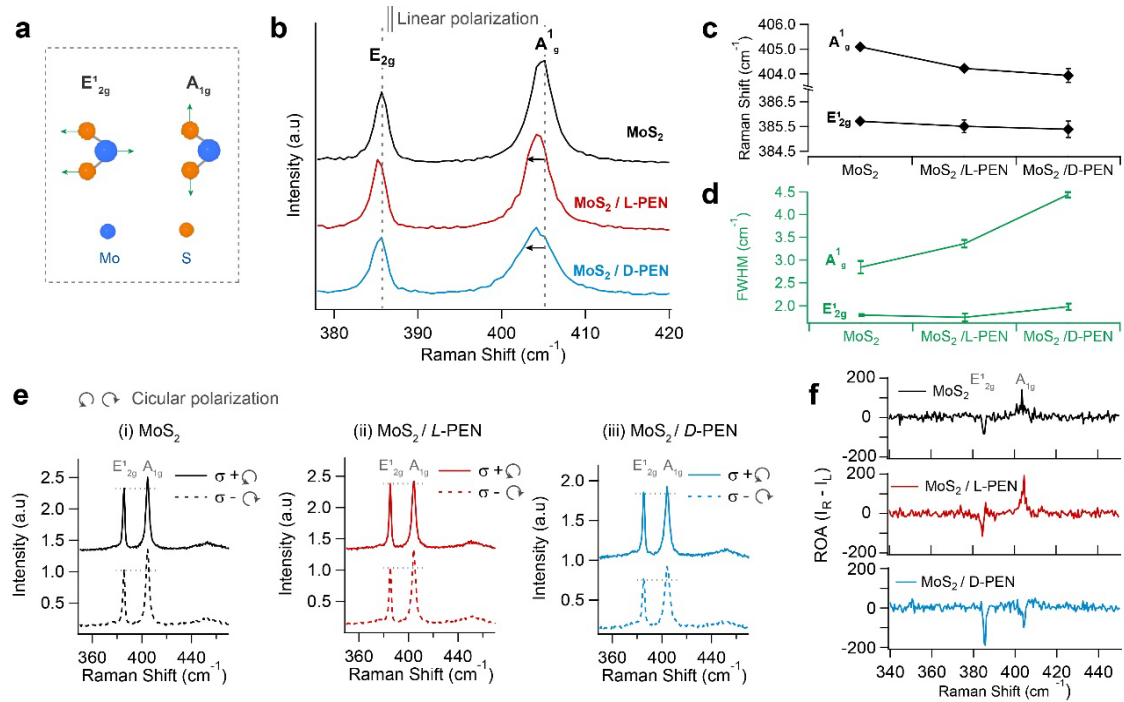

**Figure 3.** (a) Schematic representation of E$^1_{2g}$ and A$_{1g}$ Raman vibrational modes of monolayer MoS$_2$. (b) Raman spectra of MoS$_2$, MoS$_2$/ L-PEN and MoS$_2$/ D-PEN excited with 532 nm linearly polarized light. (c-d) Comparison of (c) peak position and (d) full width half maximum (FWHM) of MoS$_2$, MoS$_2$/ L-PEN and MoS$_2$/ D-PEN with 532 nm linearly polarized light. (e) Raman spectra of MoS$_2$, MoS$_2$/ L-

PEN and MoS$_2$/ D-PEN excited with 532 nm left-handed (σ+) and right-handed (σ-) circularly polarized light. (**f**) Calculated Raman optical activity (ROA) of MoS$_2$, MoS$_2$/ L-PEN and MoS$_2$/ D-PEN.

The strong chiral light-matter interactions can be used for photodetection of circularly polarized photons. As 2D MoS$_2$ is a direct band gap semiconductor with high incident photon to electron conversion efficiency (IPCE), large asymmetric absorption of left and right circularly polarised (LCP or RCP) photons by chiral 2D MoS$_2$/PEN will produce photoelectrons based on the polarization states of incident photons. 2D MoS$_2$ has been widely studied as ultrahigh-gain (>10$^3$ A/W) broad-band photodetectors (ultraviolet to infrared). However, reports of devices that show sensitivity to circularly polarized light are scarce.[82-85] We have therefore fabricated phototransistors based on chiral 2D MoS$_2$/PEN absorbers as shown schematically in **Figure 4(a)**. The gate-dependent photocurrent ($I_{ph}$) was monitored by applying a drain-source bias voltage ($V_{ds}$) and back-gate voltage ($V_g$) while illuminating the device with circularly polarized light. The incident photon energy was selected to be 3.06 eV (405 nm) which is close to the peak of CD band. Pristine MoS$_2$ shows strong photodoping as indicated by the increase in current from 10$^{-12}$ A (at $V_g$= –30 V) under dark to 10$^{-6}$ A when illuminated with 178 μW/cm$^2$ circularly polarized light. $I_{ph}$ of LCP and RCP overlap because pristine MoS$_2$ is achiral [**Figure 4(b)**]. For phototransistors with chiral MoS$_2$/L-PEN, the dark current remains at 10$^{-12}$ A but LCP illumination induces photocurrent of $0.5 \times 10^{-6}$ A, while RCP photons increase it to 10$^{-9}$ A. The giant selection of photon handedness in MoS$_2$/L-PEN is consistent with preferential of absorption of LCP over RCP light – meaning that more photons with left circular polarization participate in the excitation process to generate photoelectrons in MoS$_2$ as is also evidenced in output curves ($V_{ds}$-$I_{ds}$) in **Figure S8** in the Supporting Information. We evaluate the efficiency of incident photon to electron conversion of by calculating the photoresponsivity (*R*):

$$R = \frac{I_{ph}}{P \times A} \qquad (5)$$

Where *P* and *A* are incident light power and active photodetection area, respectively. Gate-dependent *R* is plotted in **Figure S9** in **Supplementary Information**. *R* of both pristine MoS$_2$ and chiral MoS$_2$/L-PEN reaches a maximum value of 10$^2$ A/W, which is typical for CVD-grown monolayer TMDs.[86-88] While for chiral MoS$_2$/L-PEN, *R* of LCP surpasses that of RCP by two orders of magnitude – demonstrating large photon to electron conversion anisotropy. The photodetection anisotropy factor (*g* factor) is defined by:

$$g = \frac{2\times(R_{LCP}-R_{RCP})}{R_{LCP}+R_{RCP}} \quad (5)$$

where $R_{LCP}$ and $R_{RCP}$ stands for the $R$ of left- and right- circularly polarized light. In the best performing devices, we measured a $g$ factor of 1.98 at $-30$ V, which is close to the theoretical upper limit of 2.0 (**Figure S10** in the **Supplementary Information**) – signifying that the chiral states of photons are fully distinguishable. Average $g$ factor of numerous devices is reported in **Figure 4(c)**. Negative $V_g$ leads to higher $g$ factor of up to $1.33\pm0.37$ while positive $V_g$ decreases the $g$ factor to $0.64\pm0.32$. The selective photodetection behaviour is also robust – exhibiting reliable cyclability [**Figure 4(d)**]. For $MoS_2$/D-PEN devices, we observed similar selectivity, albeit with lower photocurrent as shown in **Figure S11** in the Supporting Information. The light power dependent measurements reveal that the chiral photoresponse is possibly coming from an enhanced/supressed photogating effect under LCP/RCP illumination, which is analysed in detail in **Figure S12-S13** and the **Supporting note 1** in the **Supplementary Information**. We therefore conclude that chiral $MoS_2$/L-PEN devices are highly selective, sensitive, and responsive photodetectors for circularly polarized light compared to devices from 3D materials despite $MoS_2$/L-PEN being only 1 nm thick **Figure 4(e)**].

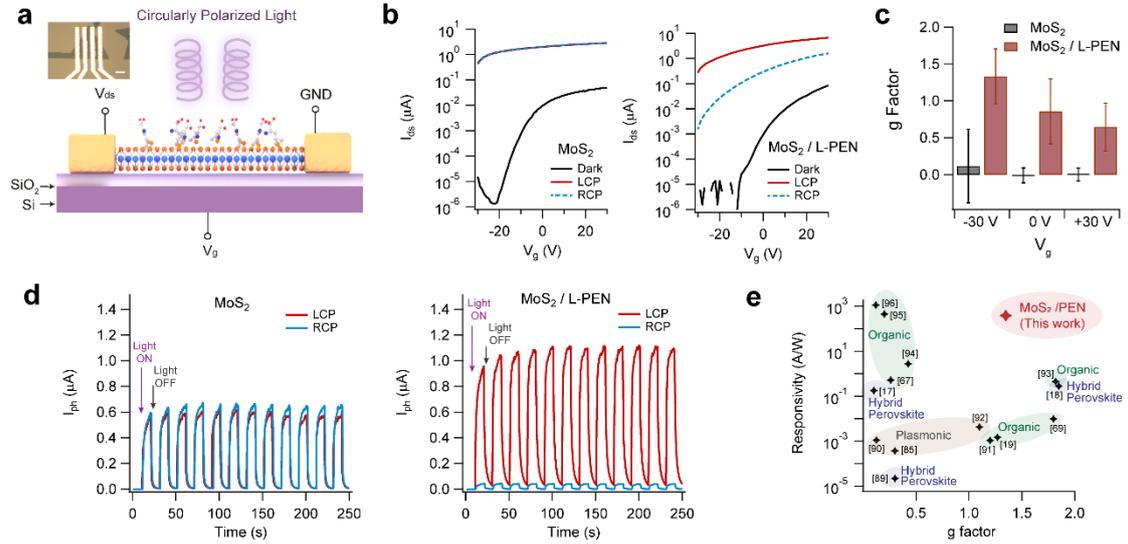

**Figure 4. (a)** Device structure of phototransistor for circularly polarized light detection. The inset shows optical microscope photo of a typical measured device. The scale bar is 10 μm. **(b)** Gate-dependent photoresponse $MoS_2$ and $MoS_2$/ L-PEN illuminated by left and right circularly polarized light (LCP and RCP). **(c)** Calculated $g$ factor of photoresponsivity of $MoS_2$ and $MoS_2$/L-PEN phototransistor under different gate bias. **(d)** Time-dependent photoresponse of $MoS_2$ and $MoS_2$/L-PEN under $V_g$= 0V and $V_{ds}$=1V of bias. **(e)** Comparison of photoresponsivity as a function of g-factor of $MoS_2$/PEN with reported values [17-19,67,69,85,89-96].

3. CONCLUSIONS

In summary, we demonstrate chirality transfer in 2D $MoS_2$ by chemical functionalization with chiral thiol-based molecules. The sub-nanometre thick functionalised monolayer exhibits ultrahigh molar ellipticity mediated by the chemisorbed chiral thiol. The chiral molecule imparts local atomic chiral field that influences the interactions between the 2D TMD and incident photons. The strong chirality-selective light-matter interactions give rise to regulation of the photoinduced electrons in the semiconducting $MoS_2$. The high molar ellipticity in chemically activated chiral 2D $MoS_2$ can be used for highly efficient detection of circularly polarized photons.